# I-GWAS: Privacy-Preserving Interdependent Genome-Wide Association Studies[*]

Túlio Pascoal[†]
University of Luxembourg
Luxembourg

Jérémie Decouchant[†]
Delft University of Technology
The Netherlands

Antoine Boutet
University of Lyon, INSA Lyon, Inria CITI
France

Marcus Völp
University of Luxembourg
Luxembourg

## ABSTRACT

Genome-wide Association Studies (GWASes) identify genomic variations that are statistically associated with a trait, such as a disease, in a group of individuals. Unfortunately, careless sharing of GWAS statistics might give rise to privacy attacks. Several works attempted to reconcile secure processing with privacy-preserving releases of GWASes. However, we highlight that these approaches remain vulnerable if GWASes utilize overlapping sets of individuals and genomic variations. In such conditions, we show that even when relying on state-of-the-art techniques for protecting releases, an adversary could reconstruct the genomic variations of up to 28.6% of participants, and that the released statistics of up to 92.3% of the genomic variations would enable membership inference attacks. We introduce I-GWAS, a novel framework that securely computes and releases the results of multiple possibly interdependent GWASes. I-GWAS continuously releases privacy-preserving and noise-free GWAS results as new genomes become available.

## 1 INTRODUCTION

Genome-wide association studies (GWASes) aim at identifying genotype/phenotype associations – the genomic variants that are statistically associated with a certain observable trait, such as a disease – in a cohort of individuals. In a GWAS, genomes of the case population – a group of individuals who express the trait of interest – are compared to a control population comprised of individuals not expressing this trait. Promises are enormous, from early prediction and prevention of diseases to personalized medicine, and are of primary interest to multiple stakeholders, including governments, research, and private companies.

To improve the confidence and precision of results, biocenters join forces and perform federated GWASes [34, 64, 67]. In particular, the possibility to share genomic data among several independent institutions, possibly located in different countries, leads to much more accurate, precise, and hence much more expressive statistical results than would be obtained by just considering the individually possessed datasets [19, 68]. Additionally, open data access accelerates medical/health research findings [7, 47]. Therefore, GWAS results should be made public whenever safely possible [53, 64].

Unfortunately, inconsiderately releasing GWAS results gives rise to privacy attacks [11, 40, 46, 48, 82]. Sankararaman et al. [70] and Zhou et al. [88] detailed conditions that enable the safe release of GWAS. The former work is based on statistical inference methods to measure the probability of inferring membership of participants using likelihood-ratio tests (LR-tests), while the latter argues in terms of the algorithmic complexity of recovery attacks.

Several works proposed privacy-preserving schemes to secure the computation of a GWAS in federated environments [8, 17, 19, 75], but they do not protect the release of their results. These works are also limited to a static setting, where the genomic dataset remains constant with exactly one GWAS release over the data. Recently, DyPS [64] considered a scenario where the results of a single-GWAS gets updated as more genomes become available or when genomes are removed (a requirement imposed by privacy regulations, such as GDPR [62]).

However, in real-life settings, GWASes might also consider overlapping sets of individuals and genomic variations, some of which might also be re-used in other studies for economic reasons [21, 46, 61, 82]. Indeed, it is rather likely that federations will run different GWASes simultaneously (e.g., one on diabetes and a second studying lung cancer) that possibly share data [27, 63]. In this work, we show that an adversary might be able to base an attack on the observation of the results of a single multi-trait study or even combining results from multiple GWASes of a federation. Besides, to the best of our knowledge, existing federated GWAS protocols cannot prevent these attacks (as shown in Table 6).

We theoretically establish and experimentally confirm that both naïve and individually-safe releases enable privacy attacks in such multi-GWAS settings. For instance, we show that an adversary exploiting as little as two overlapping GWASes, can learn about the participation of involved individuals by performing a membership attack if no further protection is enforced. Moreover, an adversary can reconstruct genomic variations of up to 28.6% of the participants by launching a recovery attack leveraging overlapping data, even if individually each release of a study is safe.

To address these issues, we introduce I-GWAS, a privacy-aware solution for releasing to the public the results of interdependent and dynamically updated Genome-Wide Association Studies. I-GWAS successfully prevents privacy risks from such interdependent GWASes by withholding from studies only those genomic variations that would enable membership inference attacks, and by ensuring that the solution space is large enough to make recovery attacks intractable, compensating in particular the reduction of this search space due to overlapping data sets. The selection of which variants should be withheld is based on a statistical inference method that bounds the risk of membership attacks [71], which I-GWAS extends to overlapping studies.

The present work makes the following contributions:

---

- We quantify theoretically and experimentally the additional privacy risk of recovery and membership attacks introduced by interdependent GWASes.
- Leveraging existing TEE-enabled federated GWAS architecture models, we propose I-GWAS, a novel privacy-protection mechanism that enables privacy-preserving and dynamic interdependent releases of GWASes.
- We extensively evaluate I-GWAS in several scenarios and against Differential Privacy (DP) based schemes for dynamic releases of interdependent GWASes.

Our analysis is based on real datasets comprised of 27,895 genomes. Although both DP and I-GWAS successfully protect interdependent GWAS results, we found our solution to exhibit a better utility-privacy trade-off, in particular as the number of releases increases. While DP inevitably has to stop releasing data once the privacy budget is exhausted, I-GWAS' exhaustive verification method ensures a continuous release of unperturbed GWAS statistics over a variable number of genomic variations.

The rest of this paper is organized as follows. Sec. 2 provides background on genomics, GWAS-associated privacy attacks, and existing countermeasures. Section 3 presents our assumptions and our objectives. Sec. 4 explains how interdependent GWASes facilitate recovery attacks. Sec. 5 presents our framework, I-GWAS, which takes interdependencies into account to enforce privacy. Sec. 6 discusses membership attacks on interdependent GWASes, while Sec. 7 shows how I-GWAS prevents them. Sec. 8 presents our experimental evaluation. Finally, Sec. 9 discusses the related work and Sec. 10 concludes this paper.

## 2 BACKGROUND ON GWAS

Humans share 99.9% of their genetic code. Possible differences at precise locations are called *genomic variations*. Short Nucleotide Polymorphisms (SNPs) are the most typical variations and describe the mutation of a single nucleotide. The most commonly found nucleotide in a population is the *major allele*, whereas the rarest is the *minor allele*.

GWASes analyze the SNPs of case and control groups to determine the significance of a given SNP for the investigated trait. Table 1 provides an example with $N$ haploid genome sequences $g_n \in \{g_1, \ldots, g_N\}$, analyzed over $L$ variants $\{SNP_1, \cdots, SNP_L\}$. Note that we consider haplotype sequences, i.e., single-strand, as in [64, 70, 71, 88]. For each SNP in this record, a 1 encodes the fact that a genome contains a minor allele, while a 0 denotes its absence. The phenotype column details whether the individual expresses the analyzed trait (e.g., diabetes), and therefore if he/she belongs to the case or control group. Results are expressed as aggregates computed from the information in the individual columns (in this example, single allele counts).

### 2.1 GWAS aggregate and test statistics

To identify the SNPs that are frequently associated with a phenotype, a GWAS aggregates genomic data into Minor Allele Frequencies (MAF), single and pairwise allele frequencies, Linkage Disequilibrium (LD), or $\chi^2$ tests [6]. These metrics allow the identification of the top alleles whose frequencies differ the most between the case and control groups. To compute those statistics, allele contingency tables that contain the counts of the possible alleles of a SNP in each population are computed.

**Table 2: A singlewise contingency table for phenotype $p$.**

|  | Phenotype$_p$ | | |
|---|---|---|---|
|  | Population | | |
| SNP$_l$ | Case | Control | Total |
| 0 (major) | $N_M^{case}$ | $N_M^{control}$ | $N_M$ |
| 1 (minor) | $N_m^{case}$ | $N_m^{control}$ | $N_m$ |
| Total | $N^{case}$ | $N^{control}$ | |

Table 2 illustrates the contingency table associated to a variant SNP$_l$ and a certain phenotype $p$. $N_i^{pop}$ is the count of major/minor alleles $i \in \{M, m\}$ in the case/control population $pop \in \{case, control\}$. $N^{case}$ and $N^{control}$ are the size of the case and the control population, respectively. $N_M$ and $N_m$ are the overall counts of major and minor alleles, respectively. Similarly, Table 3 illustrates pairwise allele frequencies of two SNPs, SNP$_i$ and SNP$_j$. $C_{ij}^{pop}$ reports the number of occurrences of the four possible combinations of alleles $\{00, 01, 10, 11\}$ in a population.

**Table 3: A GWAS pairwise contingency table for two variants. SNP$_i$ and SNP$_j$, where $i, j \in \{1, \cdots, L\}$.**

|  | Phenotype$_p$ | | | | | |
|---|---|---|---|---|---|---|
|  | SNP$_j$ | | | SNP$_j$ | | |
| SNP$_i$ | 0 | 1 | Total | 0 | 1 | Total |
| 0 | $C_{00}^{case}$ | $C_{01}^{case}$ | $C_{0-}^{case}$ | $C_{00}^{control}$ | $C_{01}^{control}$ | $C_{0-}^{control}$ |
| 1 | $C_{10}^{case}$ | $C_{11}^{case}$ | $C_{1-}^{case}$ | $C_{10}^{control}$ | $C_{11}^{control}$ | $C_{1-}^{control}$ |
| Total | $C_{-0}^{case}$ | $C_{-1}^{case}$ | $2N^{case}$ | $C_{-0}^{control}$ | $C_{-1}^{control}$ | $2N^{control}$ |

The LD test identifies a correlation among any two SNP$_i$ and SNP$_j \in \{1, ..., L\}$ and can be measured from the *p-value* on $r^2 = \frac{(C_{00}^{i,j} \cdot C_{11}^{i,j} - C_{01}^{i,j} \cdot C_{10}^{i,j})^2}{C_{0-}^{i,j} \cdot C_{1-}^{i,j} \cdot C_{-0}^{i,j} \cdot C_{-1}^{i,j}}$. On the other hand, the $\chi^2$ hypothesis test determines whether or not to reject the null hypothesis, which states that allele frequencies in the case and control populations follow a similar distribution. The $\chi^2$ statistic of a single SNP is defined as $\chi^2 = \sum_{i \in \{0,1\}} \frac{(N_i^{case} - N_i^{control})^2}{N_i^{control}}$. From the $\chi^2$ statistics, one can compute the *p-value* of each SNP. A *p-value* lower than a given threshold (e.g., $10^{-8}$) indicates that the variant might be significant [6].

Integrating genomic data from several data holders by combining GWASes over different phenotypes gives more expressive insights [21, 46, 61]. For such multi-GWAS analyses, several contingency tables such as the above ones are generated across individual analyses, and their genotype-phenotype correlations are individually computed for interpreting GWASes findings.

**Table 1: GWAS genome encoding.**

|  | SNP$_1$ | SNP$_2$ | ... | SNP$_L$ | Phenotype |
|---|---|---|---|---|---|
| Genome $g_1$ | 0 | 1 |  | 1 | Case |
| Genome $g_2$ | 1 | 1 |  | 1 | Control |
| ⋮ | | | | | |
| Genome $g_N$ | 0 | 1 |  | 0 | Case |
| Result | $N_m^{case}/N_M^{control}$ | | | | 0 (major) |
|  | $N_M^{case}/N_m^{control}$ | ... | | ... | 1 (minor) |

## 2.2 Privacy attacks on GWAS results

The results of a GWAS are typically communicated along the set $L$ of SNPs and the set $N$ of genome pseudonyms considered in the study. Based on the observation of GWAS results and metadata, researchers have reported two types of privacy attacks.

**Recovery attacks** aim at reconstructing the allele sequences of individuals (i.e., inferring unknown cells of Table 1) [88]. Knowing a set of alleles of a given genome, adversaries may proceed with identifying the participation of individuals in a GWAS [70, 88].

**Membership attacks** aim at determining whether an individual whose genome is known belongs to a specific population [40, 71]. By doing so, an adversary can associate the victim with the studied phenotype. This has profound privacy implications (e.g., adversaries discover that the victim has the investigated disease). Such illicitly obtained information can lead to unfair treatment (e.g., different insurance conditions, discrimination, etc.).

## 2.3 Safe release of GWAS results

To the best of our knowledge, our work is the first to establish conditions for safe releases of statistics from multiple GWASes that might share genomes and/or SNPs. In the following, we briefly discuss how our methods relate to recent works.

*Recovery attacks* on allele frequencies and test statistics are NP-hard [88]. Despite that, Zhou et al. argue that one needs to consider a GWAS release safe only if the solution space an adversary has to explore is significantly larger than the GWAS result space [88]. Let us introduce Zhou et al.'s safety condition for a single GWAS study comprised of $L$ SNPs and $N$ genomes. The size of the *solution space* $|S|$ is the number of possible matrices that verify a given statistical result. $|S|$ is at least $\binom{2^L}{N}$, that is, the complexity of selecting $N$ SNP sequences from $2^L$ sequences into which the $L$ SNPs expand. The size of the allele *frequency space* $|D|$ is equal to $(N+1)^{L+\binom{L}{2}}$, which corresponds to all possible values for $L$ single SNPs and $\binom{L}{2}$ SNP pairs over $N$ sequences. The condition for a safe GWAS release of this single study is that the size of the solution space $S$ is larger than the size of the frequency space $D$, i.e., $|S|>|D|$, which is equivalent to:

$$\frac{(2N-1)}{log(N+1)} > L \quad (1)$$

In this work, we show that an adversary can reduce its search space during a recovery attack if GWASes share data, and therefore possibly succeed in an attack. Some works have studied how the presence of more complex correlations in genomic data might compromise privacy. Such correlations include dependent records within a given study, e.g., the presence of individuals' relatives [1–3, 43–45]), or higher-order SNP correlations [29, 69, 80], which might allow the inference of hidden SNPs. In contrast, our paper is the first to evaluate the privacy risks that interdependent GWASes introduce under continuous releases.

Solutions to mitigate *membership attacks* include statistical inference methods, such as the likelihood-ratio test (LR-test) to measure membership inference risk of individuals from published GWAS statistics [11, 40, 48, 70, 82] and approaches that perturb a GWAS result to enforce Differential Privacy at the expense of a reduced data utility [49, 73, 74, 76, 78, 87].

**Differential Privacy.** DP guarantees that adding or removing a record in a dataset does not substantially modify the distribution of the outcome of a query [30]. There are several means to achieve DP depending on the type of perturbation added. The most common approach uses the Laplace distribution [30, 31]. More formally, let $D$ and $D'$ be two neighboring datasets that differ by a single element, and let $O$ be the set of all possible outputs of a query. A release $R$ is $\epsilon$-differentially private if:

$$Pr[(R(D) \in O] \leq exp(\epsilon) \times Pr[R(D') \in S] \quad (2)$$

where $\epsilon$ is the privacy parameter that determines the level of privacy protection that comes as a random noise added to the outputs in $O$. When leveraging the Laplace mechanism with $l_1$ sensitivity level of a function $f$ defined as $\Delta_f = \max_{D,D'} ||f(D) - f(D')||$ (which portraits the largest change in $f$ when a single record is replaced) [32], the applied noise is derived from the Laplace distribution with mean 0 and scale $\frac{\Delta_f}{\epsilon}$. In particular, as the probabilities differ by a factor of $\epsilon$, DP has a privacy and data utility trade-off. Intuitively, a smaller $\epsilon$ means stronger privacy with lower accuracy [31].

To avoid introducing noise, the scientific community has been developing statistical methods that do not rely on data perturbation to create private releases [73]. SecureGenome (SG) [70, 71] bounds the membership inference risk of individuals from GWAS statistics releases [3, 37, 64]. We extend this line of thought and build on several previous works [40, 70, 71, 88]. In particular, we extend SecureGenome [71] privacy-protection mechanism that encompass several genomic privacy measurements.

**SecureGenome foundations**. From GWAS statistics, an adversary might infer the participation of certain individuals since the probability to detect particular genomes is a function of the size of the cohort and the number of SNPs [40, 71, 88]. To bound such risks, SG acknowledges features of the human genetic model to measure and identify privacy risks from the observation of GWAS releases [70, 82, 88]. In particular, SG applies several genome privacy verifications in combination with a LR-test to determine which SNP statistics of a given dataset can be safely released.

As a first step, SG makes sure that only independent SNPs are considered in the LR-test. To do so, SG first (i) checks SNP positions with rare allele frequencies (e.g., MAF below a given threshold, e.g., ≤ 0.05) and preliminary removes them from the original SNP-set of a study; and then (ii) discards SNPs that presents a certain level of pairwise linkage disequilibrium (e.g., *p-value* on $r^2$ below $< 10^{-5}$). Posteriorly, SG proceeds with the following LR-test analysis.

SG's LR-test null hypothesis represents the possibility for an individual of interest to belong to the study population (consisting of $N$ independent individuals) under Hardy-Weinberg equilibrium (HWE)[1]. The alternative hypothesis represents the possibility for the individual of interest merged with $N-1$ individuals (obtained as under the null hypothesis and HWE), to belong to the case population. In summary, the null hypothesis tests whether the individual is not present in both populations (case and controls) of the study. On the other hand, the alternative hypothesis measures whether the tested individual belongs to the case population. These hypotheses yield the following LR-formula (adapted from [71]):

$$LR = \sum_{l=1}^{L} [x_{n,l} \log \frac{\hat{p}_l}{p_l} + (1 - x_{n,l}) \log \frac{1 - \hat{p}_l}{1 - p_l}] \quad (3)$$

where $L$ is the number of pre-selected independent SNPs in a study (recall steps (i) and (ii) discussed above), $x_{n,l}$ is the allele information at SNP position $l$ of individual $n \in [0, N-1]$, $p_l$ is the allele frequency of SNP position $l$ in the population, and $\hat{p}_l$ is the frequency of SNP position $l$ in the reference set. According to Sankararaman et al. [71], the SG's LR-test approximates the Gaussian distribution that is parameterized by the relationship between sample size $N$, number of independent SNPs $L$, statistical power $1 - \beta$, and type I error probability (or significance level) $\alpha$ when both $L$ and $N$ are moderately large, according to the central limit theorem. The extended version of SG [71] further demonstrates that this approximation also holds when $N$ is not assumed to be large using the Lindeberg-Feller central limit theorem. Besides, the authors empirically show that this approximation also holds for small $N$ and $L$ values. We present additional details of SG's analysis in Appendix A.

---
[1]HWE states that genomic variations are stable among generations given the absence of disturbing/external factors.

As a summary, SG empirically finds a safe subset of SNPs from the original set $L$ whose statistics (if released) would keep the identification power for each individual below a specified identification power threshold. Under this approximation, the LR-test is proven to provide a higher bound on the identification power $(1 - \beta)$ of any test given parameters $(N, L, \alpha)$, according to the Neyman-Pearson lemma [71]. Additionally, SG can also cope with the existence of relatives in a cohort by enabling the use of the $\gamma$ parameter to represent the probability of detecting relatives of a particular individual. For instance, the authors found that identification power for first-order and second-order relatives is lower, and decreases with the number of exposed SNPs [3, 71].

**Comparison between SG and DP.** DP is a generic approach to protect data releases against membership inference, and therefore has also been used to protect GWAS releases [5, 34, 67, 73]. DP has the advantage of not requiring a reference set (as SG does) to generate private GWAS releases. Besides, DP does not make any assumption on the background knowledge of adversaries and its computational cost is cheap once the privacy budget is calibrated (in case of static setting). However, Liu et al. [56] have identified that the presence of correlated data within a dataset can be exploited by adversaries to breach DP guarantees. Similarly, Almadhoun et al. [1, 2] recently showed that this issue also impacts genomic privacy. In particular, they demonstrated that inference attacks might become possible when adversaries leverage dependencies (e.g., statistically linked genomic variations or kinship) among the genomes in a study. In contrast, SG impedes inference attacks by preventing the release of statistics over SNPs that have a low frequency, are highly associated with another SNP (linkage disequilibrium), or whose release would lead to a high identification power through a LR-test.

In this work, we extend SG because it does not manage a privacy budget, which would eventually be exhausted and prevent releases. However, since we consider a dynamic setting, where data is released (queried) multiple times, and the presence of overlapping studies, which might impact the privacy guarantees of both SG and DP, we identify new privacy conditions that need to be enforced.

In particular, managing the privacy budget of DP mechanisms for protecting continuous observation [13, 30] and growing databases [23] are at an early development stage, and we are unaware of a practical implementation of these concepts. Equivalently, SG does not directly support dynamic releases and overlapping datasets [64]. I-GWAS builds on SG [70] and Zhou et al's. work [88], and presents new privacy conditions for the continuous release of overlapping GWASes, whose results are continuously updated as additional genomes become available or are removed. Given the availability of the genome sequences participating in the studies, I-GWAS securely executes the LR-test leveraging the selected data (genome and SNPs) to empirically check the identification power an adversary would achieve from the observation of statistics over the selected SNPs for every particular individual. In Section 8, we provide detailed discussion on the privacy properties and guarantees that I-GWAS and a DP-based mechanism achieve under dynamic and interdependent GWASes scenario.

### 2.4 Secure and privacy-preserving processing of GWAS

A federation of biocenters aiming to conduct a GWAS could rely on several privacy-preserving schemes, such as Secure Multiparty Computation (SMC) [17, 19, 38, 75], Homomorphic Encryption (HE) [8, 83, 85], local DP-based approaches [20, 58], and Trusted Execution Environment (TEE)-based solutions [9, 12, 15, 52, 64, 68]. The works mentioned above ensure privacy while computing the results of a GWAS. However, they are insufficient to prevent privacy leaks once the GWAS result is made accessible to an adversary.

Inspired by existing TEE-based deployments for federated GWAS [9, 12, 15, 52, 68], we also opted for TEE-based implementation to compute GWAS statistics and to assess whether results can be safely released [64], but also to demonstrate that our algorithm can run in a secure environment, even if this is hosted by an untrustworthy third party. We have chosen Intel SGX [60] as a vehicle for our implementation without relying on any specific feature of SGX. Our choice for SGX is motivated by previous works leveraging this technology and by its increased availability in cloud services. However, our solution applies equally well to other TEE implementations and to secure servers deployed specifically for the task of identifying safe releases [9, 64, 68]. We present existing SGX-based architectures for federated GWAS in Table 6.

**Limitations of Intel SGX.** Despite recent works leveraging advantages of Intel SGX for GWAS [9, 12, 15, 52, 64, 68] or read alignment [33, 54], SGX-based solutions need to cope with some limitations. First, SGX has a reduced size of enclave memory (only 96 MB is available to applications executing inside the enclave). Since SGX2, the enclave memory can be expanded using dynamic memory management and paging mechanisms within enclaves [14, 15, 60]. Nevertheless, such mechanisms possibly decrease performance. Intel might remove such limitations in the future and some processors already support enclaves that manipulate up to 512 GB of memory [55]. Another problem remains related to enclave-side paging. In fact, several works have shown that SGX enclaves are vulnerable to side-channel attacks that observe and exploit the memory cache access patterns of algorithms that run inside enclaves. Recently, some countermeasures have been proposed [59], such as adapting genomic workflow algorithms to work in a data-oblivious fashion (ensuring random memory access patterns), further increasing the performance penalty one has to pay when offloading computation to enclaves. Other ad-hoc solutions have been proposed, such as encoding techniques to rearrange genomic data in a certain way so that paging attacks cannot succeed. For instance, by fitting data within 4 KB page-sized blocks [14] or by processing a limited number of SNPs at a time [15]. Finally, enclaves can also be subject to Denial-of-Service (DoS) attacks [16, 77]. Although those attacks do not compromise privacy, they might disrupt the pipeline and the expected behavior of the application. However, addressing this vulnerability falls out of the scope of this work.

## 3 SYSTEM AND THREAT MODEL

**System model.** We mirror the system model and workflow of existing TEE-based architectures for federated GWAS [9, 12, 15, 52, 64, 68]. However, assuming the presence of multiple interdependent studies. In particular, we assume a federation comprised of multiple biocenters, each sequencing genomes and requesting the addition or removal of individuals from the federation. The federation jointly operates several GWASes, where each GWAS on the set $G$ {$GWAS_1, \cdots, GWAS_G$} represents a study with a corresponding phenotype, set of SNPs and genomes, possibly added or removed dynamically to update results continuously.

Following previous works [40, 64, 71, 82, 88], we assume that all GWASes publicly release GWAS metadata, allele frequencies and test statistic introduced in Section 2. Therefore, I-GWAS determines the minimal number of genomes each GWAS should use for any type of statistics to prevent recovery attacks and performs additional checks to detect vulnerable data to avoid membership attacks. All requests are evaluated in rounds, and I-GWAS certifies that a safe batch of requests (that meets the interdependent GWASes criteria) can be selected. If a safe batch of genomes cannot be found in a round, I-GWAS aborts the round. Eventually, a safe release will take place as new genome requests come over time.

Each biocenter must ensure the genome privacy of individuals it has sequenced. For that reason, they encrypt all sequenced genomes and offload their processing into a mutually TEE-enabled server. I-GWAS performs its privacy-protection analysis inside a TEE enclave. Moreover, we assume that no information leaves the TEE before I-GWAS explicitly releases it and that cryptographic primitives are secure. In addition, TEE's remote attestation [22, 60] ensures to biocenters that the piece of software deployed in the TEE is the expected one. The encryption is constructed in such a way that only the TEE can decrypt the data from the biocenter (e.g., by placing the private key required for encrypting this biocenter's genomes into the sealed memory of the TEE [22]). Sealed memory only grants access to attested enclaves to the private data they have sealed. Furthermore, data sealing is also used to increase the scalability of TEE-based solutions by allowing the retrieval of data at later stages for future steps of an algorithm running inside an enclave [9, 22, 64].

**Threat model.** We assume that all biocenters in the federation follow the protocol and provide high-precision data [64, 67, 68, 86]. We assume that our system faces an external probabilistic polynomial time adversary capable of observing released GWASes results, which it uses to mount recovery and membership attacks following previous works [64, 70, 88].

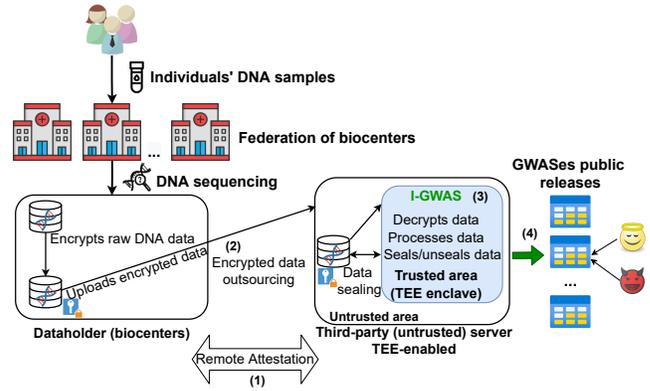

**Figure 1: A typical federated GWAS setting leveraging TEE.**

**Workflow overview.** In a round, I-GWAS proceeds in the following steps, depicted in Fig. 1: (1) Before a biocenter starts interacting with the TEE, it remotely attests the authenticity of the hardware, software, and configuration of the TEE and finally establishes a secure connection with it. (2) Each biocenter locally encrypts and transfers data to the TEE, i.e., new genomes and their corresponding requests to add genomes to a study or to remove their participation from existing studies. (3) Upon reception of requests, the TEE decrypts the genomes, includes them in the data structures it maintains, and selects a batch of genomes (ideally including the newly added genomes) that can be safely used for a candidate release while impeding recovery attacks. Next, I-GWAS performs its extended LR-test analysis to evaluate the feasibility of membership attacks on the selected genome set merged with potentially overlapping releases. Only SNPs that would not allow such an attack will be considered in the study. (4) Finally, the TEE uses the data (genomes and SNPs) selected in step (3) to compute and release the actual GWAS statistics. The TEE periodically executes this workflow.

In summary, this paper contributes precise conditions under which interdependent GWAS preserve the privacy of the individuals who share their genetic data.

## 4 SAFETY CONDITIONS FOR INTERDEPENDENT GWASES

The overlapping of SNPs and genomes can be leveraged by adversaries to reduce the solution space for inferring the matrices that verify an observed statistical result. In this section, we review recovery attacks and extend the safety conditions proposed by Zhou et al. [88] for interdependent GWASes.

In Fig. 2, we consider two studies $GWAS_1$ and $GWAS_2$ that release statistics over $L_1$ SNPs and $N_1$ genomes, and over $L_2$ SNPs and $N_2$ genomes, respectively. The studies overlap in $N_{ovl}$ genomes and $L_{ovl}$ SNPs. Individually, both studies fulfill Zhou's safety condition, that is, $|S_1|>|D_1|$ and $|S_2|>|D_2|$ (recall § 2.3).

However, leveraging knowledge about the overlapping regions of $GWAS_1$ and $GWAS_2$, adversaries might be able to reduce the search space analysis for each possible situation in which these studies may overlap (i.e., evaluating addition, subtraction and union mappings over releases' solution spaces). Eventually, if the solution space of a combination of releases is not large enough (i.e., $|D| \approx |S|$ [88] – recall § 2.3) such a combination might be subject to a recovery attack. Due to space constraints, we refer the reader to Appendix B for further details on how adversaries might map releases.

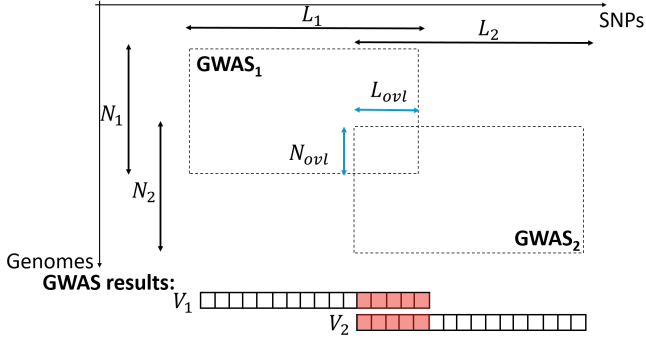

Figure 2: Illustration of two overlapping GWASes.

Next, we formally define the complexity of the search space analysis when adding, subtracting, and taking the union of statistical results for single and pairwise allele frequencies, and for test statistics when releases can be combined and leveraged by adversaries to circumvent privacy conditions in place.

### 4.1 Singlewise allele frequencies

In this case, $GWAS_1$ and $GWAS_2$ release two maps $m_1$ and $m_2$ that associate a SNP to its minor allele counts ($V_1$, $V_2$) in $N_1$ and $N_2$. The space of possible solutions for addition is therefore $|S_{add}| = 2^{L_1 \cdot N_1 + L_2 \cdot N_2 - L_{ovl} \cdot N_{ovl}}$ and $|S_{sub}| = 2^{L_1 \cdot N_1 + L_2 \cdot N_2 - 2 \cdot L_{ovl} \cdot N_{ovl}}$ for subtraction. The latter formula comes from the fact that values in the intersection of the two matrices are canceled out under subtraction. Intuitively, $|S_{union}| = |S_{add}|$.

For the frequency spaces, we obtain $|D_{add}|$ and $|D_{sub}|$ by computing the product of the number of possible values for each SNP that they contain:
$|D_{add}| = (N_1+1)^{L_1-L_{ovl}} \cdot (N_2+1)^{L_2-L_{ovl}} \cdot (N_1+N_2-N_{ovl}+1)^{L_{ovl}}$.
$|D_{sub}| = (N_1+1)^{L_1-L_{ovl}} \cdot (N_2+1)^{L_2-L_{ovl}} \cdot (N_1+N_2-2N_{ovl}+1)^{L_{ovl}}$.

$|D_{union}|$ is the product of the frequency spaces of both releases divided by the frequency space over the overlapped area: $|D_{union}| = \frac{(N_1+1)^{L_1} \cdot (N_2+1)^{L_2}}{(N_{ovl}+1)^{L_{ovl}}}$.

### 4.2 Pairwise allele frequencies

This analysis produces maps (also presented in Appendix B) that associate tuples $(i, j, p, q)$ to pairwise allele counts in the respective datasets. Here, $i$, $j$ denote the SNPs and $p$, $q$ the allele types [88].

The solution space complexities for $|S_{add}|$ and $|S_{union}|$ are the same as for singlewise allele frequencies. In contrast, $|S_{sub}| = 2^{L_1 \cdot N_1 + L_2 \cdot N_2 - (L_{ovl} \cdot N_{ovl})}$, because $m_{sub}$ depends on the values in the intersection (i.e., $(i,j) \in (N_{ovl}, L_{ovl})$), which are canceled out.

Like in the singlewise frequencies case, $|D_{add}|$ and $|D_{sub}|$ are obtained by computing the product of possible values of the SNPs over the released frequencies.

Let $|D_{N_1}| = (N_1+1)^{\binom{L_1-L_{ovl}}{2}+(L_1-L_{ovl}) \cdot L_{ovl}+(L_1-L_{ovl})}$ and $|D_{N_2}| = (N_2+1)^{\binom{L_2-L_{ovl}}{2}+(L_2-L_{ovl}) \cdot L_{ovl}+(L_2-L_{ovl})}$. Then,

$|D_{add}| = |D_{N_1}| \cdot |D_{N_2}| \cdot (N_1+N_2-N_{ovl}+1)^{L_{ovl}+\binom{L_{ovl}}{2}}$.

$|D_{sub}| = |D_{N_1}| \cdot |D_{N_2}| \cdot (N_1+N_2-2N_{ovl}+1)^{L_{ovl}+\binom{L_{ovl}}{2}}$.

$|D_{union}|$ is the product of the frequency spaces divided by the frequency space of the overlapped area:

$|D_{union}| = \frac{(N_1+1)^{L_1+\binom{L_1}{2}} \cdot (N_2+1)^{L_2+\binom{L_2}{2}}}{(N_{ovl}+1)^{L_{ovl}+\binom{L_{ovl}}{2}}}$.

### 4.3 Test statistics

From a GWAS release, apart from observing the sets of SNPs $L$ and genomes $N$ that participated in a study, adversaries can also observe the *p-values* statistics of the $\chi^2$ test along with the $r^2$ values of linkage disequilibrium. However, the $r^2$ values encompass fewer information than pairwise frequency statistics from the adversary's perspective [82, 88], which leads to the safety condition $|R^2| = \frac{(N+1)^{L+\binom{L}{2}}}{2^{\binom{L}{2}}}$ being smaller than $|D| = (N+1)^{L+\binom{L}{2}}$ (recall § 2.3 analysis).

We can therefore derive the solution space for test statistics using the same approach as for pairwise frequency space (which is actually the theoretical upper bound since other allele frequencies, such as singlewise frequencies, can be derived from them): $|S_{add}| = |S_{union}| = 2^{L_1 \cdot N_1 + L_2 \cdot N_2 - L_{ovl} \cdot N_{ovl}}$ and $|S_{sub}| = 2^{L_1 \cdot N_1 + L_2 \cdot N_2 - (L_{ovl} \cdot N_{ovl})}$.

Test statistics space complexities are derived from the product of possible values for the SNPs over the released test statistics with $|D_{N_1}|, |D_{N_2}|$ as above: $|R^2_{add}| = \frac{|D_{N_1}|}{2^{\binom{L_1-L_{ovl}}{2}+(L_1-L_{ovl}) \cdot L_{ovl}}} \cdot$

$\frac{|D_{N_2}|}{2^{\binom{L_2-L_{ovl}}{2}+(L_2-L_{ovl}) \cdot L_{ovl}}} \cdot \frac{(N_1+N_2-N_{ovl}+1)^{L_{ovl}+\binom{L_{ovl}}{2}}}{2^{\binom{L_{ovl}}{2}}}$.

$|R^2_{sub}| = \frac{|D_{N_1}|}{2^{\binom{L_1-L_{ovl}}{2}+(L_1-L_{ovl}) \cdot L_{ovl}}} \cdot \frac{|D_{N_2}|}{2^{\binom{L_2-L_{ovl}}{2}+(L_2-L_{ovl}) \cdot L_{ovl}}}$

$\cdot \frac{(N_1+N_2-2N_{ovl}+1)^{L_{ovl}+\binom{L_{ovl}}{2}}}{2^{\binom{L_{ovl}}{2}}}$.

$|R^2_{union}|$ is computed as the product of the test statistics spaces of both releases over the overlapping area:

$|R^2_{union}| = \frac{\frac{(N_1+1)^{L_1+\binom{L_1}{2}}}{2^{\binom{L_1}{2}}} \cdot \frac{(N_2+1)^{L_2+\binom{L_2}{2}}}{2^{\binom{L_2}{2}}}}{\frac{(N_{ovl}+1)^{L_{ovl}+\binom{L_{ovl}}{2}}}{2^{\binom{L_{ovl}}{2}}}}$.

In addition, although inferring correct values of $r^2$ (to measure associations between SNPs that can be used to facilitate attacks) from GWAS test statistics is NP-hard [88], I-GWAS not only conducts the searching space analysis (assuming the full SNP-set $L$) but also certifies that SNPs found to be in linkage disequilibrium (LD) do not have their statistics released. This conservative approach impedes adversaries from leveraging LD to mount attacks.

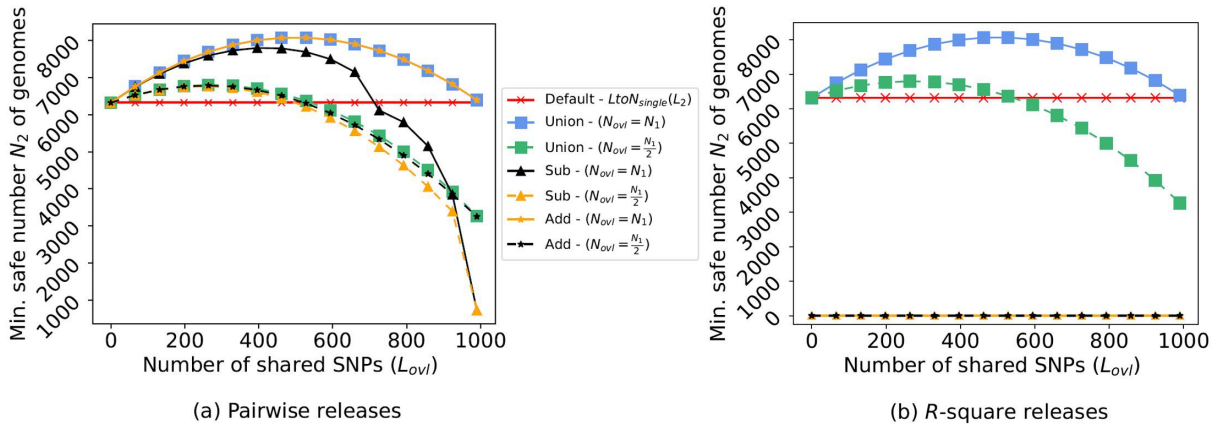

Figure 3: Smallest number of genomes $N_2$ that a GWAS that overlaps with a previous GWAS should use for a safe release depending on their overlapping SNP-set size ($L_{ovl}$) and genomes set size ($N_{ovl}$).

## 5 PROTECTING INTERDEPENDENT GWASES AGAINST RECOVERY ATTACKS

As shown in Section 4, releases of interdependent GWASes must satisfy particular safety bounds. In particular, these new bounds enforce that the conditions of Eq. 1 also holds for overlapping releases so that the solution space of combinations of releases is sufficiently large. We now detail how I-GWAS protects the release of statistics against recovery attacks using sequential releases, which assumes that studies are dynamically updated as new genomes are sequenced and/or removed [64]. We start by providing the intuition behind the release of GWAS statistics using an example where only one interdependent GWAS has been previously released, before generalizing to $G$ GWASes.

### 5.1 Sequential releases of GWASes

Let us assume that a first GWAS$_1$ has released statistics over $L_1$ SNPs with $N_1$ genomes, and that a second, GWAS$_2$, aims at releasing statistics over $L_2$ SNPs. We note $LtoN_{single}(L)$ the minimum number of genomes to use to release the results of a single GWAS on $L$ SNPs (cf. Equation 1). Our approach consists in increasing the number $N_2$ of genomes that GWAS$_2$ uses so that the safety bounds for interdependent studies are verified. Particularly, we discovered that releasing GWAS$_2$ with $LtoN_{single}(L_2)$ genomes is not safe with overlapping datasets.

We evaluate in Fig. 3 the smallest safe value for $N_2$ when $L_1$= 1,000 and $N_1=LtoN_{single}(L_1)$, and when $L_{ovl} \in [0, L_1]$ and $N_{ovl} \in \{\frac{N_1}{2}, N_1\}$. In this figure, the default line represents $LtoN_{single}(L_2)$, i.e., the state-of-the-art formula that assumes independent and single releases. By observing the behavior of $N_2$, one can notice that depending on the attack (i.e., targeting the addition, subtraction, or union), $N_2$ exceeds $LtoN_{single}(L_2)$. In such situations, we can identify that interdependent releases are not safe if relying on $LtoN_{single}(L_2)$. In addition, it can be noticed that protecting against the union and addition attack always requires more genomes than for the other cases.

Besides, we can note that the addition and subtraction attack lines against $r^2$ releases (Fig. 3 (b)) always stay below the default line (note that the lines plotted for the addition and subtraction mappings overlap each other at the bottom of the chart). This means that adding or subtracting $r^2$ values to launch a privacy attack is not practical and matches previous works findings [82, 88].

Interestingly, one can also notice that $N_2$ decreases when $L_{ovl}$ increases for all type of attacks. This downwards behavior comes from the fact that the solution spaces (e.g., $|S_{add}|$) grow faster than the frequency spaces (e.g., $|D_{add}|$) with $L_{ovl}$, and because a recovery attack is deemed possible depending on their ratio (e.g., $|S_{add}|/|D_{add}|$) [88].

In summary, interdependent releases are safe when the space analysis for each type of attack is kept within safe boundaries, i.e., the combined solution space between the releases is sufficiently larger than their combined frequency spaces. In particular, the number $N_2$ of genomes required to protect interdependent releases depends on $L_{ovl}$ and $N_{ovl}$. For instance, for these experiments, if such conditions are not enforced, up 28.6% of the genomes are vulnerable to recovery attack.

Therefore, to mitigate recovery attacks, I-GWAS identifies $N_2$ such that $|S_{op}|>|D_{op}|$ for $op \in \{add, sub, union\}$ given the overlaps a study has with previous GWAS. For pairwise allele frequencies and for test statistics, the union attack provides the theoretical bound for interdependent releases. Hence, it is sufficient to check that $N_2$ verifies $|S_{union}|>|D_{union}|$. For singlewise allele frequencies (not illustrated in Fig. 3 for space reasons), the subtraction attack defines the safety bound. In summary, I-GWAS selects the largest bound as the safety threshold and apply its conditions when selecting safe batches of genomes for the creation of safe releases.

### 5.2 Scaling with the number of GWASes

We now extend this method to the case where $G$ GWASes have previously released statistics. Let us note $|S_i|$ the solution space for a given GWAS$_i$, and $|D_i|$ its frequency set space. Due to space restriction, we present only the analysis for the pairwise frequency space. The analysis for the other statistics (i.e., singlewise and $r^2$) is similar.

Before introducing I-GWAS' solution for multiple releases of interdependent GWASes, let us first discuss an intuitive solution to this problem. Indeed, to enforce that several interdependent releases are safe, one could compute all possible combinations of existing releases. Then, measure and evaluate if the solution space is sufficiently large given the released statistics space. Such a brute force approach is secure but has exponential complexity and is not reasonable in practice [64]. Motivated by that, I-GWAS offers a novel solution with linear complexity. I-GWAS relies on the following Theorem that defines how and when multiple interdependent GWASes can be dynamically released without infringing on the genomic privacy of their participants.

THEOREM 1 (SAFE RELEASES OF INTERDEPENDENT GWASES). *For every $GWAS_i$ and $GWAS_j$, if (1) $|S_i|>|D_i|$ and (2) $\frac{|S_j|}{|D_j|} > \prod_{i \neq j} |S_i \cap S_j|$, then any combination of releases from $GWAS_i$ and $GWAS_j$ involves enough genomes to prevent recovery attacks.*

PROOF. The very first release of a GWAS in the federation does not overlap with any other GWAS, and therefore meets the condition for independent GWAS releases, i.e., $|S_i|>|D_i|$. The release is safe. Let us assume that any combination of $i \geq 1$ GWAS releases is safe. Let further $j$ be the id of the $(i+1)$-th GWAS, $\{S_1, S_2, ..., S_i\}$ the sets of the $i$ previous GWAS solution spaces, and $\{D_1, D_2, ..., D_i\}$ their corresponding frequency spaces. All these GWASes also contain enough genomes to meet the single-GWAS safe condition (i.e., $|S_j|>|D_j|$). The inclusion-exclusion formula states that

$$\left| \bigcup_{j=1}^{i} E_j \right| = \frac{\prod_{j=1}^{g} |E_j|}{\frac{\prod_{1 \leq j < k \leq g} |E_j \cap E_k|}{\frac{\prod_{1 \leq j < k < l \leq g} |E_j \cap E_k \cap E_l|}{\cdots}}}{(-1)^{g-1}|E_j \cap \cdots \cap E_g|}$$

where all $|E_j|$ can be substituted by either $|S_j|$ or $|D_j|$ to compute the sizes of the combination solution and frequency spaces, respectively. Given this formula, one easily obtains that $|\bigcup_{j=1}^{i} S_j| \geq \frac{\prod_{j=1}^{g} |S_j|}{\prod_{1 \leq j < k \leq g} |S_j \cap S_k|}$ and that $1 \leq |\bigcup_{j=1}^{i} D_j| \leq \prod_{j=1}^{g} |D_j|$. Therefore, if one ensures that $\frac{\frac{\prod_{j=1}^{g} |S_j|}{\prod_{1 \leq j < k \leq g} |S_j \cap S_k|}}{\prod_{j=1}^{g} |D_j|} \geq 1$ then $|\bigcup_{j=1}^{i} S_j| > |\bigcup_{j=1}^{i} D_j|$. This condition is equivalent to $\prod_{j \leq g-1} \frac{|S_g|}{|D_g|} \cdot \left( \frac{|S_j|}{|D_j|} \cdot \frac{1}{\prod_{k \leq g} |S_g \cap S_k|} \right) \geq 1$, which is provided since $\prod_{j \leq g-1} \frac{|S_g|}{|D_g|} \geq 1$ because of condition (1), and since $\frac{|S_j|}{|D_j|} \cdot \frac{1}{\prod_{k \leq g} |S_g \cap S_k|} \geq 1$ because of condition (2). □

I-GWAS relies on Theorem 1 to verify that a GWAS can release or update its results, given that other GWASes have already released theirs. This verification has a complexity that is linear with the number of GWASes. To illustrate this process, we present its pseudocode in Appendix C as Algorithm 1.

## 5.3 Allowing safe genome removals

I-GWAS provides dynamic processing of genomes and their safe removal, similarly to DyPS [64], and extends it to interdependent GWASes. For the update of a given (single) GWAS, DyPS assembles a batch of genome additions and removals, respectively represented by $A$ and $R$, such that $|A| + |R| \geq LtoN_{single}(L)$ and $|A| \geq |R|$. The rationale behind this approach is that even if some genomes that participated in a GWAS are removed, the solution space an adversary has to explore only increases with time. I-GWAS, however, offers a method that acknowledges interdependent GWASes. Before a GWAS can release its results, I-GWAS determines the minimum number of genomes it should use, which might also depend on other GWASes and which we denote as $LtoN_{I\text{-GWAS}}$, to satisfy the conditions of Theorem 1.

Let us consider an example with two (possibly overlapping) releases of two interdependent GWASes. The first GWAS, $GWAS_i$, consists of genome additions $A_i$ and removals $R_i$, and the second $GWAS, GWAS_j$, consists of genome additions $A_j$ and removals $R_j$. Since these releases were orchestrated by I-GWAS, we have $|A_i| + |R_i| \geq LtoN_{I\text{-GWAS}}(i)$ and $|A_j| + |R_j| \geq LtoN_{I\text{-GWAS}}(j)$. Let us then assume that $GWAS_i$ is updated with new genome operations $A_{i'}$ and $R_{i'}$, such that $A_{i'} > R_{i'}$. Then, each GWAS considered alone stays safe. The number of remaining genomes in $GWAS_i$ is $|R_i| + |A_i \setminus R_{i'}| + |R_{i'} \setminus A_i| + |A_{i'}| > |R_i| + |A_i| > LtoN_{I\text{-GWAS}}(i) \geq LtoN_{single}(i)$. The combination of $GWAS_i$ and $GWAS_j$ is also safe because Theorem 1 enforces that it contains more than $LtoN_{I\text{-GWAS}}(i)$ genome operations.

# 6 MEMBERSHIP ATTACKS ON INTERDEPENDENT GWASES

As with novel recovery attacks presented in Section 4, an adversary can leverage the fact that some genomes might have been used in multiple interdependent studies for succeeding in membership attacks. In particular, an adversary can launch membership attacks over release combinations, which might increase the identification power of the attack.

DyPS [64] combines SG's LR-test (cf. §2.3) with an exhaustive verification process to dynamically update the statistics of a single GWAS. In particular, every SNP in a candidate release is checked against existing releases to identify if it statistics has been released before by another study. DyPS does not consider multiple interdependent GWASes. In practice, we show that an adversary could combine statistics across several overlapping releases and mount a successful membership attack (as shown by our experiments in Section 8). Therefore, we detail additional required verifications to support dynamic releases of interdependent GWASes.

Like previous works [37, 64, 71] I-GWAS also leverages SG for membership protection. However, in contrast to existing solutions, which evaluate the conditions of safe releases considering studies separately, I-GWAS offers a novel pipeline able to protect the privacy of participating genomes by implementing an exhaustive verification step that acknowledges all possible sets of genome and SNPs combinations among existing studies. Particularly, I-GWAS applies an exhaustive local verification (on a single-GWAS level) combined with a global verification that considers all existing combinations of GWASes on a per-SNP basis.

# 7 PROTECTING INTERDEPENDENT GWASES AGAINST MEMBERSHIP ATTACKS

After selecting a safe batch of genomes using the conditions presented in Section 4 to prevent recovery attacks on interdependent GWASes presented in Section 5, I-GWAS identifies data that can be released without posing membership privacy risks. I-GWAS leverages SG's to run membership inference tests over the selected genomes and SNPs of the candidate GWAS (recall Section 2.3). Thus, identifying the set of safe SNP positions regarding a candidate release selected data.

However, the above verification is not enough. Indeed, I-GWAS also needs to enforce that the selected SNPs statistics can be safely updated within a single GWAS previous releases (i.e., verifying which SNPs are "locally" safe). This task consists in executing additional LR-test verifications over all possible combinations of releases (and so genome distributions) within previous releases of a particular study (depicted on the left side of Fig. 4). Recall that rare alleles and SNPs in LD are primarily blocked from participation, and therefore such information could not be leveraged by adversaries.

In addition, to prevent an adversary from leveraging the combination of released statistics from other studies, I-GWAS retains in the candidate GWAS only those SNPs of the previous step that remain "globally" safe. For this, I-GWAS first identifies overlapping SNPs and then executes additional LR-tests over the combination of genomes from heterogeneous studies that shared SNPs with the candidate GWAS (depicted on the dashed lines coming from the right side of Fig. 4). After this procedure, I-GWAS has identified a list of SNPs that survived all verifications and therefore can be used for a safe release.

Let us consider the example illustrated in Fig. 4, where $GWAS_1$ releases take place first, and SNPs are selected following the local verification only (represented by solid lines) because there was only one study. The notation $N_{i\_r}$ represents the genome set selected for $GWAS_i$, release $r$. We now present and discuss the different scenarios I-GWAS considers when any candidate release is found:

**Local verifications (solid lines).** A first situation occurs when a SNP that is labeled as safe by I-GWAS and has never been studied before can be released without global verification because an adversary cannot combine releases. This is the case for all selected SNPs in $N_{1\_1}$ of $GWAS_1$ and $id_4$ in $N_{2\_1}$ of $GWAS_2$. When SNPs have been considered in previous releases of a study, e.g., $id_1$, $id_2$ and $id_3$ of $GWAS_1$ at $N_{1\_2}$ and $N_{1\_3}$, those releases are combined by I-GWAS and extra LR-tests rounds are conducted to certify that candidate SNPs are also identified as safe when the release combinations are tested. In this example, $id_1$ succeeds in all local verifications of $GWAS_1$ releases (green lines). In contrast, $id_2$ failed (red lines) when tested against the ($N_{1\_2}$, $N_{1\_1}$) combination. Thus, $id_2$ statistics are withheld in $N_{1\_2}$ release.

**Local and global verifications (dashed lines).** In this case, a SNP is found to be safe for a single GWAS release, but has been released in another study before. For example, $id_1$ is safe over the $N_{2\_1}$ genome set of $GWAS_2$, but $id_1$ statistics has been released in $GWAS_1$ before (over the overlapping genome set $N_{ovl}$). In this case, I-GWAS evaluates new LR-test rounds to certify that a SNP is also identified as safe over the combinations of genomes among the two studies. This is represented by the dashed line coming from the selected SNPs of $GWAS_2$. In our example, $id_3$ is detected as safe over the $N_{2\_1}$ set in $GWAS_2$, but when combined with the other genome sets from $GWAS_1$, the LR-test identified that this SNP cannot have its statistics safely released anymore (dashed red lines). On the other hand, $id_1$ passed all tests when $GWAS_1$ releases are combined with those of $GWAS_2$ (dashed blue lines), and therefore might have its statistics published. SNP $id_4$ is not tested since it has never been released before by any study.

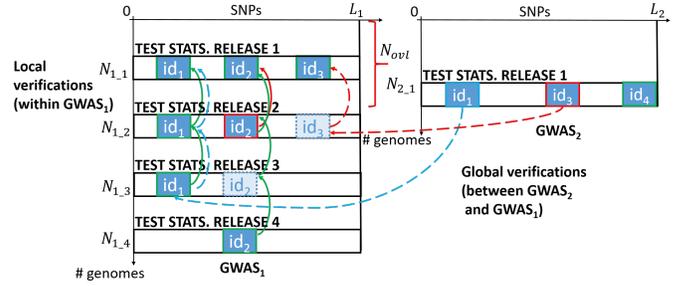

**Figure 4: Exhaustive verification process to protect interdependent GWASes releases against membership attacks.**

Furthermore, to avoid that previously released SNPs (and potentially not considered in the future) being leveraged by adversaries, I-GWAS keeps track of all released SNPs so that when combining releases for further verification, those SNPs are also checked. These "ghost" SNPs are represented by dashed boxes ($id_2$ and $id_3$) in Fig. 4, which allow I-GWAS to apply the exhaustive verification step for all released SNPs regardless if they can currently be observed or not. For instance, even though SNP $id_2$ was not released in $N_{1\_2}$ of $GWAS_1$, it is checked anyway once it is a candidate SNP selected for $N_{1\_4}$ release. In that example, $id_2$ has been labeled as safe over all runs of membership verification and therefore is allowed to have its statistics released at $N_{1\_4}$ of $GWAS_1$.

In summary, the goal of I-GWAS's exhaustive verification is to replicate the behavior of an adversary that accumulates knowledge by combining releases to launch membership attacks leveraging overlapping data. Thus, certifying that only the SNPs mutually selected as safe after multiple membership test verifications (over the combinations of overlapping releases and their respective genomes) are allowed to release statistics. As a result, maintaining the identification power of individuals within safe boundaries over all releases and therefore impeding the adversary to use such a strategy.

We present the pseudocode of I-GWAS for the membership tests in Algorithm 2 (Appendix C). The complexity of I-GWAS' verification increases with the number of releases and studies. The computational complexity for the verification is $O(L' \cdot 2^{LocalRel \cdot OverlappedRel})$, where $L'$ is the number of selected SNPs after the first run of the LR-test on the candidate release set, $LocalRel$ is the number of releases within a study, and $OverlappedRel$ is the number of overlapping releases from other GWASes. The full I-GWAS' workflow for interdependent GWASes is shown in Algorithm 3.

## 8 EXPERIMENTAL EVALUATION

We implemented and run I-GWAS on Intel SGX using C++ and Graphene SGX [77]. Experiments were performed on an Intel i7-8650U machine (2.11 GHz, 16GB RAM) with Ubuntu 18.04. We used real genomes from the phs001039.v1.p1 dbGAP dataset [81], which consists of 14,860 case and 13,035 control genomes. We adopt the parameters suggested in SecureGenome [71] for the genomic privacy protection tests to decide which SNPs might have their allele frequencies safely released (recall §2.3). Namely, 0.05 MAF cut-off, $10^{-5}$ LD cut-off, 0.1 false-positive rate and a 0.9 detection power rate. We evaluate I-GWAS along 3 metrics: *privacy*, *data utility (accuracy loss and release utility score)* and *run-time*.

We compare I-GWAS against DyPS [64], a state-of-the-art approach to dynamically release GWAS privacy-preserving results, and against $\epsilon$-DP using Laplace mechanism for the dynamic releases of GWAS statistics. We consider several $\epsilon$ and privacy budgets (*pvb*) spent over releases to evaluate trade-offs. *pvb* is used to keep DP properties over multiple releases. *pvb* starts at 1 (100% of $\epsilon$) and each release consumes a fraction of $\epsilon$. When *pvb* is exhausted, DP cannot release data with original privacy guarantees. We utilized PyDP, a Python wrapper for Google's Differential Privacy C++ library [36].

We do not compare I-GWAS with a method based on local DP since it introduces higher perturbation than a centralized DP scheme [20, 58]. Moreover, we identified that the system-side performance (e.g., bandwidth, memory, and CPU consumption) of I-GWAS is very similar to that of other TEE-based solutions [9, 12, 15, 52, 64, 68]. Additionally, I-GWAS only imposes a penalty of requiring extra genomes to protect overlapping releases (cf. Table 5) when compared to DP-based releases. Such a limitation depends on the assumed workload, e.g., the rate at which new genomes are added and the frequency of overlapping data (cf. Fig. 3).

To be fair when comparing I-GWAS against DP-based releases, we create a release utility metric that acknowledges both data coverage (amount of data allowed to be used in a release) and accuracy loss. Otherwise, DP utility score would always perform worse than I-GWAS. We evaluate the utility of a release as follows:

$$\sum_{l=0}^{L} \frac{\text{SNP}_{l_{rel}} \cdot \text{SNP}_{l_{acc}}}{L} \quad (4)$$

In this formula, $L$ is the original SNP-set of the GWAS, $\text{SNP}_{l_{rel}} \in \{0, 1\}$, i.e., 1 if statistics over SNP $l$ has been released and 0, otherwise, and $\text{SNP}_{l_{acc}}$ is the accuracy of the released statistics of SNP $l$ compared to its original (noise-free) result. Note that we still evaluate I-GWAS along other traditional metrics mentioned before individually.

**Privacy and data utility.** Fig. 5 illustrates the impact of privacy, i.e., preventing the release of GWAS results at some SNP positions, on data utility. In this experiment, we use all 14,860 case genomes and consider two GWASes over 10,000 SNPs and varied the fraction of overlapping genomes among studies between 1% to 50%. The vulnerable SNPs are positions that would put participating genomes at risk of being identified in a membership attack. These SNPs need to be identified and secluded from public releases. The GWAS results are only known by the I-GWAS's trusted enclave, and only the results of SNPs that can be safely exposed are publicly shared.

With the dbGAP dataset that we consider, the release coverage decreases when the number of overlapping genomes increases because more SNPs are withheld by I-GWAS. In the worst case, statistics computed over 81.8% of the SNPs could be used to identify individuals and are therefore withheld. Interestingly, in this use case, we noticed that all SNPs could be released if very few genomes are shared between studies (between 1% – 10%).

Table 4 presents the results of our second scenario where we considered 4 GWASes. The first 3 GWASes used disjoint sets of 4,953 genomes each, while the last one shares each of its 14,860 genomes with the other three GWASes. For each experiment/line, we considered a different number of SNPs. We identify that using DyPS, which cannot protect releases of interdependent studies, the number of SNP positions at risk increases with the overall number of SNPs, from 80% to 92.3%, which is aligned with the findings of Simmons et al. [73]. As a result, I-GWAS presents a smaller release coverage since it refrains statistics of vulnerable SNPs from being released. In this experiment, I-GWAS released statistics over 20% to 7.7% of the original SNP-set depending on the scenario.

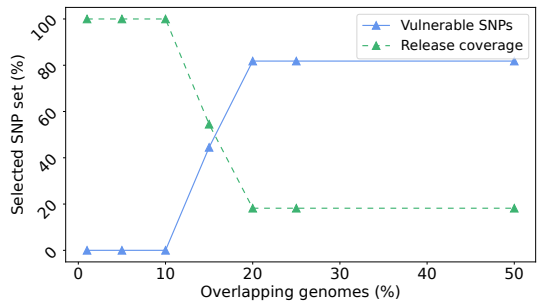

**Figure 5: Release coverage when protecting interdependent GWASes against membership attacks.**

**Table 4: I-GWAS: protection of vulnerable SNPs.**

| # SNPs | Vulnerable SNPs using DyPS [64] (%) | Release coverage w/ I-GWAS (%) |
|---|---|---|
| 1,000 | 80 | 20 |
| 2,500 | 81.8 | 18.2 |
| 5,000 | 84.6 | 15.4 |
| 10,000 | 92.3 | 7.7 |

**Comparison with $\epsilon$-DP.** We now consider a scenario where a first GWAS (GWAS$_1$) has already released allele frequencies over 1,000 SNPs using 7,430 genomes (note that $LtoN(1, 000) = 6,320$). A second study (GWAS$_2$) aims at releasing allele frequencies over 1,000 SNPs from 14,860 genomes. GWAS$_1$ shares half of the SNPs and genomes with GWAS$_2$. Therefore, $N_{ovl} = 7,430$ and $L_{ovl} = 500$. We compare the overall score of the release up to the GWAS$_2$ release, using I-GWAS or DP. Setting $\epsilon$ and a privacy budget (*pvb*) with DP is necessary to support a given number of releases. These parameters interfere on the noise applied to protect a release [23, 31]. We repeat this experiment 100 times and report the average results. We illustrate the results of this experiment in Table 5.

I-GWAS evaluates that the second study would need 350 additional genomes (i.e., 5.53% more genomes) than the state-of-the-art $LtoN$ formula, which indicates 6,320 genomes would be enough to protect GWAS$_2$, in order to prevent recovery attacks on overlapping studies. Relying on additional genomes to enforce privacy slightly delays releases. Future work could be to use synthetic genomes while preserving statistical properties for this purpose [41, 66].

$\epsilon$-DP releases statistics over all SNPs and presents a better release utility score in several scenarios (compare blue and green scores in Table 5), but it perturbs the results (accuracy loss column). The accuracy loss metric corresponds to how distant the DP result is from the original statistics one would obtain without privacy guarantees. While I-GWAS release utility score is equal to 66 irrespective of the number of subsequent releases, DP release utility scores varied from 76.41 to 94.11 in the best cases. Nevertheless, DP showed poor performance in settings that would allow more future releases (red scores in Table 5). In fact, the limited privacy budget of DP-based releases restricts the number of conceivable releases, whereas I-GWAS can afford releases by virtue of its exhaustive verification methods. To allow more releases, $\epsilon$-DP should use smaller $pvb$ over releases so that some privacy budget is kept to protect future releases. Nevertheless, smaller $pvb$ means increased accuracy loss. Intuitively, using larger values of $pvb$ over releases would increase the release utility of $\epsilon$-DP releases but reduces the number of allowed safe releases. Therefore, when adopting DP, GWAS federations have to carefully select the privacy budget that will be spent over dynamic releases. For instance, considering $pvb = 0.12$ per release, $\epsilon$-DP could afford only up to 8 safe releases with decreased utility. In contrast, using half of the total privacy budget per release ($pvb = 0.50$) would only allow two safe releases but higher utility.

**Table 5: Comparison between I-GWAS and the standard $\epsilon$-DP using Laplace mechanism (average of 100 repetitions).**

| Approach | Maximum # releases | Accuracy loss (%) | Coverage (% SNPs) | Required additional genomes (%) | Release utility score |
|---|---|---|---|---|---|
| $\epsilon$=0.5 ($pvb$=0.12) | 8 | 290.98 | 100 | 0 | 1.86 |
| $\epsilon$=0.5 ($pvb$=0.25) | 4 | 116.46 | 100 | 0 | 52.77 |
| $\epsilon$=0.5 ($pvb$=0.33) | 3 | 85.09 | 100 | 0 | 63.85 |
| $\epsilon$=0.5 ($pvb$=0.50) | 2 | 52.62 | 100 | 0 | 76.41 |
| $\epsilon$=1 ($pvb$=0.12) | 8 | 122.15 | 100 | 0 | 50.8 |
| $\epsilon$=1 ($pvb$=0.25) | 4 | 52.48 | 100 | 0 | 76.46 |
| $\epsilon$=1 ($pvb$=0.33) | 3 | 38.29 | 100 | 0 | 82.37 |
| $\epsilon$=1 ($pvb$=0.50) | 2 | 24.95 | 100 | 0 | 88.2 |
| $\epsilon$=1.5 ($pvb$=0.12) | 8 | 75.67 | 100 | 0 | 84.32 |
| $\epsilon$=1.5 ($pvb$=0.25) | 4 | 33.90 | 100 | 0 | 88.97 |
| $\epsilon$=1.5 ($pvb$=0.33) | 3 | 25.02 | 100 | 0 | 90.66 |
| $\epsilon$=1.5 ($pvb$=0.50) | 2 | 16.52 | 100 | 0 | 92.87 |
| $\epsilon$=2 ($pvb$=0.12) | 8 | 55.95 | 100 | 0 | 86.08 |
| $\epsilon$=2 ($pvb$=0.25) | 4 | 25.15 | 100 | 0 | 90.64 |
| $\epsilon$=2 ($pvb$=0.33) | 3 | 18.78 | 100 | 0 | 92.27 |
| $\epsilon$=2 ($pvb$=0.50) | 2 | 12.33 | 100 | 0 | 94.11 |
| I-GWAS | $\infty$ | 0 | 66 | 5.53 | 66 |

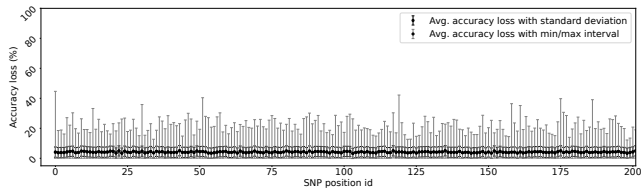

**Figure 6: Analysis of the accuracy loss per SNP using $\epsilon$-DP releases (cut-off of the first 200 of 1,000 SNPs).**

**Analyzing the impact of noise with $\epsilon$-DP.** Assuming the same scenario, we also evaluate in more detail the accuracy loss impact of $\epsilon$-DP on allele frequencies of each SNP. The error bar plots show the average (black circles, ideally at "0", i.e., without noise), standard deviation (black rectangles), minimal and maximal values (grey lines) of accuracy loss of the DP-based releases over 100 repetitions. Fig. 6 presents a cut-off of the first 200 SNP positions of the release for the second study using $\epsilon = 2$ and $pvb = 0.12$ (the setting with highest utility score and maximum releases).

Allele frequencies were applied 4.14% of noise on average, with a 0.41% average standard deviation, which kept the perturbation applied over SNPs results in the 3.73 - 4.55% accuracy loss interval. For some SNPs, original statistics were distorted above 40%, e.g., SNP id 1, 51 and 119. Using I-GWAS, the same study released statistics over 66% of the original SNP-set without any noise addition.

**Run-time and complexity.** Fig. 7 shows I-GWAS's run-time for a GWAS that overlaps with 5 previous GWASes that consider 500 (left side) or 1,000 SNPs (right side). We assumed that half of the genomes and SNPs that participated in the $G$ previous GWASes releases were re-used by the current GWAS. First, a batch of requests for the candidate release is selected individually, and then its requests are checked against previous overlapping studies. Evaluating all possible combinations of existing releases with the brute force method has an exponential complexity. However, thanks to Theorem 1, I-GWAS's run-time scale linearly with the number of existing releases and are shorter than those of a brute force approach. In addition, I-GWAS has a linear complexity even when considering a larger number of SNPs. For instance, with 1,000 SNPs, I-GWAS run-time varied from 111 seconds for 2 GWASes to 157 seconds for 5 GWASes, whereas the brute force approach lasted from 111 seconds to 955 seconds, respectively. A similar behavior also happened with the experiments over 500 SNPs. While I-GWAS varied from 4 to 14 seconds, the brute force needed 4 to 69 seconds according to the number of GWASes, respectively. For the membership inference protection, we measured the average run-time for one analysis over the largest dataset (i.e.,14,860 genomes and 10,000 SNPs). I-GWAS computes this verification in 12.73 seconds. Keeping the number of SNPs and considering fewer genomes, I-GWAS takes on average 6.45 seconds for 9,906 genomes, 4.61 seconds for 7,430 genomes, and 2.96 s for 4,953 genomes. Lastly, the average memory consumption of I-GWAS in the enclave was 2 MB. Hence, respecting SGX's memory limitations.

**Further discussion.** Both I-GWAS and DP are valid approaches and can even be complementary to each other for membership protection of GWAS releases. They should be chosen according to the expectations of the federation in terms of precision of the results, foreseen number of releases and privacy-protection guarantees. Unfortunately, due to the unavailability of DP-based mechanisms for continuous dynamic releases of GWASes, we claim that that giving up accuracy (to the scale as presented in Table 5) for privacy when using standard DP, might not be reasonable, especially when dealing with high-precision studies, such as GWAS [72, 73].

In I-GWAS, we have chosen to extend SecureGenome [71] to cope with dynamic and overlapping releases given its genome-oriented nature, for being used in previous TEE-based privacy-preserving architectures [64], and because it is a noise-free approach. We show that I-GWAS can afford an infinite number of releases thanks to its exhaustive verification scheme, which is not the case when using DP because limited privacy budget. In particular, I-GWAS certifies that individuals' identification power is kept within safe boundaries even when an attacker has the ability to observe and combine releases. Moreover, we show that I-GWAS is scalable and has linear complexity depending on the number of existing overlapping studies. However, I-GWAS's utility score might be impacted depending on the number of vulnerable SNPs found in the original SNP-set. Moreover, I-GWAS needs additional genomes to protect releases against recovery attacks, which depends on how large are overlapping regions among studies.

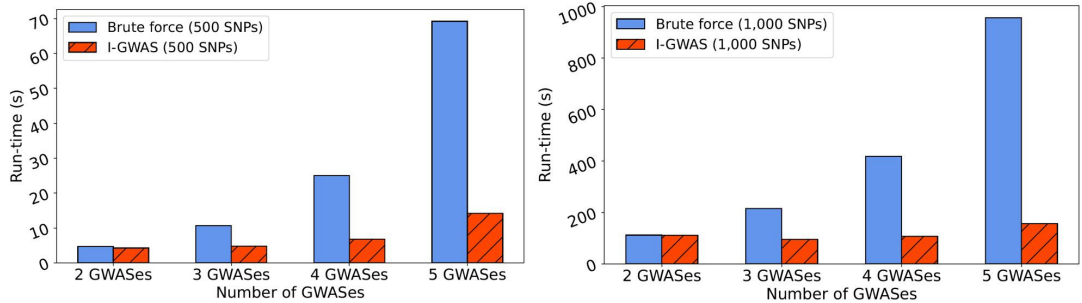

Figure 7: Run-times of the brute force approach and of I-GWAS to protect against recovery attacks on interdependent GWASes.

I-GWAS and DP might be vulnerable to the presence of dependent records [1, 2, 23, 30]. However, Sankararaman et al. [71] demonstrated that SG can also be configured to detect the presence of relatives in a cohort. I-GWAS can be extended to include this feature, while we are not aware of this possibility for DP-based solutions. Besides, we envision that genome-oriented privacy protection methods can be combined with DP mechanisms to provide better utility. In particular, running I-GWAS as a first step to detect vulnerable SNPs (which need further protection that can be enforced with DP-based releases), while the statistics of the remaining SNPs can be released without perturbation. Thus, increasing the overall utility score of releases. Such a study is left for future work. Finally, even though we evaluated I-GWAS using the same privacy parameters adopted in previous works [64, 71], we recall that I-GWAS can also support stricter privacy-protection levels by specifying more conservative thresholds and statistical confidence levels.

**I-GWAS's limitations.** I-GWAS considers SNP correlations up to the level of pairwise linkage disequilibrium (LD), like several previous works [42, 71, 76, 82, 88]. Several attacks assume adversaries that might leverage high-order correlations [29, 69, 80], phenotype data [29], and kinship [3, 28, 29, 43, 44]. Recent research on genome-wide LD identified that: (i) using $k^{th}$ Markov Chain Models (MCM) to identify higher-order correlations might not scale, whereas leveraging recombination models is linear with the number of SNPs [29]; (ii) inference power of an attack does not increase much when considering MCM with $k > 3$ [69]; (iii) relying on a Hidden Markov Model (HMM) presents better accuracy than MCM [29, 69]. I-GWAS can be extended to remove SNPs that are involved in $k^{th}$ higher-order correlations or SNPs that could be inferred using HMM. Future work would be required to determine the minimum order of correlations to consider to prevent privacy attacks under dynamic and overlapping settings. Besides that, we note that I-GWAS's exhaustive verification is computationally more expensive than DP-based algorithms. Moreover, we note that the number of SNPs I-GWAS is able to update depends on the current number of overlapping studies and how high their data are correlated, which can impact the release coverage of studies, as verified in our experiments. Lastly, although novel discoveries of correlations are becoming rarer [18], I-GWAS can adapt its algorithm to changing control population statistics but only a posteriori.

## 9 RELATED WORK

There is a known tension between genomic data sharing and privacy [50, 79] in genomics research, and a pressure to enable individuals to control how their data are used [24, 25]. To comply with current data-privacy regulations' constraints, such as the US HIPAA the EU GDPR, data subjects shall have the right to withdraw their consent to participate in a GWAS at any time. Lastly, data holders are encouraged to produce public releases of GWAS [7, 47, 64].

**Privacy-preserving GWAS.** Cho et al. [17] offer a SMC mechanism where both individuals and computing parties (CPs) privately share their data using the Beaver multiplication triples secret sharing mechanism. Constable et al. [19] propose a *Secure Two-Party Computation* (STPC) approach to perform privacy-preserving computation of $\chi^2$ and MAF. Tkachenko et al. [75] offer a similar framework to compute $\chi^2$ and $p$-tests. These approach only consider two parties, and therefore cannot be used in practical and large-scale federated GWAS. Several methods leveraged homomorphic encryption (HE) to store encrypted genomes in a cloud and compute GWAS statistics [10, 51, 57]. Aziz et al. [5] presented new DP-based algorithms that dynamically manage privacy budgets of the DP mechanism to find optimal values for $\epsilon$. MedCo [67] is a distributed protocol built on Unlynx [34], that allows exploratory medical analysis combining HE, DP, and other improvements to compute statistics over medical data privately. SAFETY [68] combines HE for the aggregating data holder's genomic data inputs and SGX enclaves for more complex statistical processing. Similarly, SCOTCH [16] uses SGX and HE to gather genomic data from several data holders and compute aggregate statistics in a faster manner. More recently, Bomai et al. [9] combined multi-key HE and SGX to enable secure sharing of genomic data from multiple data holders to a SGX-enabled cloud provider that computes $\chi^2$ GWAS statistics to answer queries from authorized users.

These works focus on the sharing and processing part, and not on the privacy-preserving release of GWAS. If publicly released, their results might be subject to genomic privacy attacks. In contrast, *I-GWAS enforces privacy during both data processing and releasing of interdependent GWASes leveraging a TEE-based framework and privacy-preserving release algorithms.*

**Privacy-preserving release of GWAS.** Cai et al. [11] describe a practical membership attack based on as little as 25 random SNPs, while He et al. [39] leverage belief propagation methods. Humbert et al. [43] show that an attacker might correctly infer genotype information of individuals by leveraging statistical relationships among genomic variants. Ayday et al. [3] identified a victim's hidden genomic data by using data of its relatives. DP has been used to enforce safe releases of GWAS. Uhlerop et al. [78] release GWAS statistics over the $M$ most significant SNPs using Laplace noise. Jiang et al. [49] propose to use a new privacy-budget approach that balances data perturbation with privacy risks using statistics of a LR-test. DP was also used to enable differentially private logistic regression by perturbing the objective function [84] instead of the final output of a GWAS. Simmons et al. [72] leverage the neighbor distance algorithm to apply noise more efficiently. PrivSTRAT applies data perturbation considering the group stratification in a study, and PrivLMM is based on Linear Mixed Models (LLMs) [74]. In a later study, Simmons et al. [73] enable differentially private releases of GWAS relying on Bayesian statistics, MCM and HMM. Furthermore, Almadhoun et al. [1, 2] have shown that the existence of dependent records in a genomic database decrease the privacy guarantees of DP mechanisms. Several works aimed at providing DP in a dynamic environment, such as DP under continual observation [13, 30] and for growing databases [23]. To the best of our knowledge, these techniques have not been used under a dynamic GWAS or closely related scenario. Recently, Ayoz et al. [4] showed that recovery and membership attacks can be launched by sequentially querying genomic data-sharing Beacons.

Alternative methods ensure that the identification power over released genomes is sufficiently low without data perturbation. For this purpose, one can use the hypothesis test with $T_r$ [82] or $\Lambda$ [88] metrics. Recently, GenDPR [65] proposes a distributed implementation of the LR-test that only manipulates intermediary results in a TEE and does not require genomic data to be exchanged.

*We establish conditions for safe interdependent GWASes under continuous releases, and propose algorithms to enforce them.* I-GWAS extends and combines genome-oriented privacy-preserving release algorithms with exhaustive verification methods to provide genomic privacy over continuous releases without data perturbation.

**Overview of federated GWAS solutions.** Table 6 investigates recent federated GWAS works and their properties. In summary, few works [5, 64, 67] combine privacy-preserving processing with privacy-preserving releasing of GWAS. Only DyPS and I-GWAS support public GWAS results releases that are dynamically updated. More importantly, I-GWAS is the only solution to allow dynamic privacy-preserving releases of interdependent GWASes.

## 10 CONCLUSION

We highlighted that sharing genomes and SNP positions in different GWASes requires the development of new privacy-preserving methods. In those settings, we showed that previous techniques that aim at preventing recovery and membership attacks might be ineffective. Using known statistical methods that enable safe and dynamic data releases for a single independent GWAS, we evaluated that up to 28.6% of the processed genomic information could be recovered by an adversary in a scenario that involved two interdependent GWASes. We also identified that between 80% and 92.3% of the SNPs might expose individuals to membership attacks.

To fill this gap, we present I-GWAS, a novel framework that extends privacy-preserving algorithms to protect the genomic privacy of individuals participating in interdependent GWASes. I-GWAS offers a linearly scalable algorithm to select which genomes and SNPs can safely participate in releases of GWAS statistics, while not allowing recovery and membership attacks even if adversaries can leverage overlapping data among releases.

I-GWAS does not introduce noise and produces safe releases by slightly delaying the release of statistics until sufficiently enough additional genomes become available and withholding vulnerable SNPs. In contrast, although Differential Privacy can be adapted to support dynamic releases of overlapping studies and sometimes presents a better release utility score, it limits the number of possible releases and the accuracy of the released statistics.

Future work include analyzing and improving the interplay between differential privacy and tolerable privacy budgets, and combining differential privacy with I-GWAS to improve its utility.

Table 6: Overview of federated GWAS solutions. GC: Garbled circuit; *: Releases *yes/no* answers, which is vulnerable to similar Beacon's privacy attacks [4].

| Work | Cryptographic scheme | Privacy-preserving releases | Public releases | Dynamic releases | Interdepend. releases |
|---|---|---|---|---|---|
| Cho et al. [17] | SMC | ✗ | ✗ | ✗ | ✗ |
| Tkachenko et al. [75] | SMC (ABY framework [26]) | ✗ | ✗ | ✗ | ✗ |
| Hasan et al. [38] | HE + GC-based SMC | ✗ | ✗ | ✗ | ✗ |
| Sadat et al. (SAFETY) [68] | TEE (Intel SGX) | ✗ | ✗ | ✗ | ✗ |
| Carpov and Tortech [12] | TEE (Intel SGX) | ✗ | ✗ | ✗ | ✗ |
| Kockan et al. [52] | TEE (Intel SGX) | ✗ | ✗ | ✗ | ✗ |
| Bomai et al. [9] | TEE (Intel SGX) | ✗ | ✗ | ✗ | ✗ |
| Blatt et al. [8] | HE | ✗ | ✗ | ✗ | ✗ |
| Froelicher et al. [35] | Multiparty HE | ✗ | ✗ | ✗ | ✗ |
| Bonte et al. [10] | HE and SMC | ✓* | ✗ | ✗ | ✗ |
| Aziz et al. [5] | Improved $\epsilon$-DP | ✓ | ✗ | ✗ | ✗ |
| Raisaro et al. [67] | HE + DP + Unlynx [34] | ✓ | ✗ | ✗ | ✗ |
| Pascoal et al. [64] | TEE (Intel SGX) | ✓ | ✓ | ✓ | ✗ |
| This work (I-GWAS) | TEE (Intel SGX) | ✓ | ✓ | ✓ | ✓ |

## A  SECUREGENOME - DETAILS

Likelihood-Ratio tests (LR-tests) have been used to evaluate the probability for individuals to be reidentified given released GWAS statistics [40, 82]. Both privacy attacks and membership inference protection mechanisms used LR-tests [11, 64, 70, 88]. SecureGenome (SG) [70, 71] performs a LR-test and several additional statistical checks to bound the privacy risks for each individual participating in released GWAS statistics.

**Goal.** The goal of SG is to select a subset of SNPs from the original SNP-set of a GWAS over which statistics can be computed and released without allowing any membership inference. To reach that goal, SG applies several checks using the pool of genomes participating in a study and a reference genome set, which is assumed to be available.

**Assumptions.** SG assumes that the SNPs considered in the LR-test are independent, i.e., in linkage equilibrium, which can be justified when selected SNPs are located in distant regions. In addition, SG assumes that there are no genotyping errors, i.e., the allele information over the SNPs are precise. This is a common assumption in the literature. However, SG's authors have shown by experimentation that genotyping errors decrease the identification power of the attack. Furthermore, SG's LR-test assumes that SNP allele frequencies in the population are strictly bounded by 0 and 1. Hence, there is an $a > 0$ such that $a \leq p_l \leq 1 - a$, where $p_l$ corresponds to the allele frequency of SNP $l$ in the cohort. This is an expected assumption due to the fact that GWAS only considers SNPs whose minor allele frequencies (MAF) are well represented in the selected population.

SecureGenome executes the following processing steps:

**Step 1: Removing SNPs with rare allele frequencies.** In this step, SG pools and computes the allele frequencies at each position (considering both case and control genome sets) and checks if the MAF of SNPs are below or equal to a parameter $MAF_{\text{cutoff}}$, usually equal to 0.05. SNP positions with low MAF are outliers that could be used by adversaries to infer membership. They are therefore not considered for the subsequent steps.

**Step 2: Removing SNPs in high LD.** High linkage disequilibrium between two SNPs (e.g., p-value on $r^2 < 10^{-5}$) can be leveraged to attack individuals by using the association levels among SNPs, as shown in [70, 88]. For this step, SG employs a greedy algorithm to remove one of two SNPs that are in LD. If any two SNP positions are found to be in LD, the most ranked SNP is kept for further verification. Notice that Steps 1 and 2 enforce the assumption that SNPs are independent for the subsequent LR-test analysis.

**Step 3: Identifying SNPs that would allow membership inference.** A LR-test verification is performed over the remaining SNPs at specified detection power threshold and false-positive rates. SG uses the remaining SNPs from the previous steps to conduct the LR-test described below.

SG's null hypothesis draws the probability of an individual belonging to the case genome set consisting of $N$ genomes under Hardy-Weinberg Equilibrium (HWE) coming independently from the distribution drawn from the reference genome set. In contrast, SG's alternative hypothesis draws the probability of the individual belonging to the pool consisting of $N - 1$ genomes under the null hypothesis and HWE. In summary, the null hypothesis tests whether the individual is not present in any population of the study (case and control genomes). On the other hand, the alternative hypothesis measures whether the tested individual belongs to the case population. These hypotheses yield the formula presented in Equation 3 of Section 2.3.

The power $1 - \beta$ of the LR-test can be found as a function of the pool size $N$, the number of SNPs $L$ with a tolerable false-positive rate $\alpha$. Conversely, using the Neyman-Pearson lemma that states that no test can have larger power than the LR-test, the power $1 - \beta$ achievable for the LR-test given $(L, N, \alpha)$ determines the largest $L$ so that no $(\alpha, \beta)$-test can be obtained for a pool of size $N$.

In [71], Sankararaman et al. demonstrate that SG's LR-test approximates the Gaussian distribution that is parameterized by the relationship between sample size $N$, the number of independent SNPs $L$, statistical power $1-\beta$, and type I error probability (or significance level) $\alpha$ when both $L$ and $N$ are moderately large, according to the central limit theorem (cf. [71], §T8, pp. 40-43). Additionally, the authors demonstrates that the LR-test also approximates the Gaussian distribution thanks to the Lindeberg-Feller central limit theorem when $N$ is not assumed to be large (cf. [71], §T8.2, pp. 43-44)

In particular, SG uses the LR-test to empirically verify the identification power of the $N$ individuals in a study by sampling their allele sequences over several subsets of SNPs $\in L$ in an iterative fashion while removing SNPs that would keep the identification power of any participant above a specified detection power threshold. Thus, SG prevents the release of GWAS statistics of SNPs that would pose membership inference risks of any individual.

To conclude, after its LR-test, SG has identified a SNP-set $L'$ belonging to the original SNP-set $L$ (i.e., $'L \in L$), from which the observation of GWAS statistics over these SNPs would not allow an adversary to identify the presence of individuals participating in the study respecting the configured (i) upper bound on the power (the probability that an individual is correctly identified to be in the population); and (ii) upper bound on the false positive rate (the probability that an individual that is not in the population is erroneously identified to be in the population) of the LR-test.

Therefore, the SNPs belonging to $L'$ can then be safely used in a GWAS as their statistics would not allow inferring the presence of underlying individuals at the specified confidence level of the test.

## B  RECOVERY ATTACK MAPPINGS

In the following, we precise the statistics an adversary is able to compute from the results of two GWASes. From these formulas, one can verify the formulas we provided in Section 4 for $D_{add}$, $D_{sub}$, $D_{union}$, $S_{add}$, $S_{sub}$, and $S_{union}$ for allele frequencies and for test statistics.

**Singlewise statistics mapping.** From the $m_1$ and $m_2$ maps in Fig. 2 an adversary can compute three other maps, $m_{add}$, $m_{sub}$ and $m_{union}$, whose elements are defined over $L_1 \cup L_2$ as follows:

$$m_{add}[i] = \begin{cases} m_1[i] & \text{if } i \in L_1 \setminus L_2 \\ m_2[i] & \text{if } i \in L_2 \setminus L_1 \\ m_1[i] + m_2[i] & \text{if } i \in L_2 \cap L_1 \end{cases}$$

$$m_{sub}[i] = \begin{cases} m_1[i] & \text{if } i \in L_1 \setminus L_2 \\ -m_2[i] & \text{if } i \in L_2 \setminus L_1 \\ m_1[i] - m_2[i] & \text{if } i \in L_2 \cap L_1 \end{cases}$$

$$m_{union} = \{(m_1, m_2) \in \{0,1\}^{(L_1 \cdot N_1)} \cdot \{0,1\}^{(L_2 \cdot N_2)}$$
$$| m_1[i,j] = m_2[i,j] \text{ for } (i,j) \in (N_{ovl}, L_{ovl})\}$$

**Pairwise statistics mapping.** Following the same idea, from $m_1$ and $m_2$, one can compute other maps, $m_{add}$, $m_{sub}$ and $m_{union}$, whose elements are defined for $(i,j) \in N_1$ or $(i,j) \in N_2$ as follows:

$$m_{add}[i,j,p,q] = \begin{cases} m_1[i,j,p,q] & \text{if } (i,j \in L_1 \setminus L_2) \\ & \vee (i \in L_1 \setminus L_2 \wedge j \in L_{ovl}) \\ m_2[i,j,p,q] & \text{if } (i,j \in L_2 \setminus L_1) \\ & \vee (i \in L_2 \setminus L_1 \wedge j \in L_{ovl}) \\ m_1[i,j,p,q] \\ \quad + \\ m_2[i,j,p,q] & \text{if } i,j \in L_{ovl} \end{cases}$$

$$m_{sub}[i,j,p,q] = \begin{cases} m_1[i,j,p,q] & \text{if } (i,j \in L_1 \setminus L_2) \\ & \vee (i \in L_1 \setminus L_2 \wedge j \in L_{ovl}) \\ -m_2[i,j,p,q] & \text{if } (i,j \in L_2 \setminus L_1) \\ & \vee (i \in L_2 \setminus L_1 \wedge j \in L_{ovl}) \\ m_1[i,j,p,q] \\ \quad - \\ m_2[i,j,p,q] & \text{if } i,j \in L_{ovl} \end{cases}$$

$$m_{union} = \{(m_1, m_2) \in \{0,1\}^{(L_1 \cdot N_1)} \cdot \{0,1\}^{(L_2 \cdot N_2)}$$
$$| m_1[i,j,p,q] = m_2[i,j,p,q] \text{ for } (i,j,p,q) \in (N_{ovl}, L_{ovl})\}$$

## C I-GWAS ALGORITHMS (PSEUDO-CODE)

Algorithm 1 illustrates the operations that I-GWAS performs in order to select a safe batch of genome requests considering overlapping GWASes. Every time a safe batch of genome requests for a single-GWAS is found, its requests are simulated (line 5), and then the solution spaces are checked acknowledging existing potentially overlapping releases (from line 6 to 10). In other words, the algorithm checks if the conditions presented in Sections 4 and 5 hold). If the candidate release presents safe boundaries and meets the conditions, it can proceed to the next verification.

**ALGORITHM 1** Verification of a set of genome requests to prevent recovery attacks on overlapping GWAS releases.

1: **procedure** isSafe$_{I-GWAS}$($g, G, Add\_Req, Rmv\_Req$)
2:    **Input:** $g$: candidate GWAS with $Add\_Req$ and $Rmv\_Req$ genome additions and removals, respectively; $G$: set of released GWASes
3:    **Output:** set of selected genome addition and removal requests for interdependent GWASes ($LtoN_{I-GWAS}$)
4:    **Uses:** $S_i$ and $D_i$ respectively return the solution and frequency space sizes for a GWAS $i$ (or an intersection of GWASes); $copyAndApplyRequests(g, Add\_Req, Rmv\_Req)$ applies a batch of request to a copy of a GWAS and returns it.
5:    $g' = copyAndApplyRequests(g, Add\_Req, Rmv\_Req)$
6:    $tmp = \frac{S(g')}{D(g')}$
7:    **for** $i$ in $G$ **do**
8:        $tmp = tmp \cdot \frac{S_i}{D_i} \cdot \frac{1}{S(g' \cap i)}$
9:    **end for**
10:   **return** $tmp \geq 1$
11: **end procedure**

After selecting a safe batch of genomes (that would allow a safe release against recovery attacks), I-GWAS now evaluates which SNPs might have their GWAS statistics safely released without allowing membership inference. Algorithm 2 details such a process. First, I-GWAS identifies which SNPs can be released over the bath of selected genomes for the single-GWAS candidate release (line 5) using the *SNPSelection* function that represents the SecureGenome's [70] step introduced in Section 2.3, which is used to find SNPs that can be safely exposed while avoiding membership attacks. Recall that in this step, SNPs that presents rare allele frequencies are in LD are also identified and blocked from participation.

**ALGORITHM 2** Selection of SNPs to prevent membership attacks on overlapping GWAS releases.

1: **procedure** CHECKINTERDEPENDENTMEMBERSHIP($Add\_Reqs, Rmv\_Reqs, G$)
2:    **Input:** $Add\_Req$ and $Rmv\_Reqs$ of candidate study $g$ and $G$ set of released GWASes
3:    **Output:** set of selected SNPs for safe interdependent GWASes
4:    **Uses:** $AllCombinations(relsToCombine)$ creates combinations of releases in $relsToCombine$; $SNPSelection(Add\_Reqs, Rmv\_Reqs, g)$ runs the LR-test and returns the safe SNP-set for a single GWAS $g$
5:    $selected\_SNPs := SNPSelection(Add\_Reqs, Rmv\_Reqs, g)$
6:    **for** SNP$_l$ in $selected\_SNPs$ **do** // for each selected safe SNP in a single GWAS, isSafeSingleBiocList
7:        $relsToCombine := \emptyset$
8:        **for** $g$ in $G$ **do** //check each existing GWAS $g$
9:            **for** $rel$ in $g$ **do** //check each release of GWAS $g$
10:               **if** (SNP$_l$ == $rel.s$) **then** // SNP position SNP$_l$ has been released in a release $rel$
11:                  $relsToCombine.add(rel)$
12:               **end if**
13:           **end for**
14:           **for** $combRel$ in $AllCombinations(relsToCombine)$ **do**
15:               $testSet := combRel.Add\_Reqs + combRel.Rmv\_Reqs + Add\_Reqs + Rmv\_Reqs$ // merge genomes requests
16:               $checkSafeSNPs := SNPSelection(testSet)$
17:               **if** (SNP$_l$ in $checkSafeSNPs$) **then**
18:                  **continue** // this SNP can be released
19:               **else**
20:                  $safe\_SNPs.del(SNP_l)$ // this SNP cannot be released
21:               **end if**
22:           **end for**
23:       **end for**
24:   **end for**
25:   **return** $selected\_SNPs$ // set of safe SNPs for candidate release
26: **end procedure**

Nevertheless, the standard *SNPSelection* function (i.e., SecureGenome) only works in a static GWAS release setting. To enable dynamic GWASes releases under the presence of overlapping data, I-GWAS conducts additional verifications explained in Section 7 and described in the following paragraph.

I-GWAS enforces that each SNP$_l$ (from the original SNP-set $L$ of a study $g$) that is identified as safe by the *SNPSelection* function is to be tested over all possible combinations of existing releases (lines 6 to 24). In particular, I-GWAS first identifies and collects all releases where SNP$_l$ has previously participated (lines 7 to 13). Then, I-GWAS loops and run the *SNPSelection* function for each possible combination of intersected releases that used SNP$_l$ (lines 14 to 16). If SNP$_l$ is labeled as safe in all verifications (i.e., over all combinations), it means that it can be safely released. Otherwise, SNP$_l$ is withhold from the candidate release (line 20). In the end of this loop, I-GWAS has identified a list of SNPs that survived (i.e., was labeled as safe when checked against existing combinations of releases) and therefore can have their statistics safely released (line 25).

Algorithm 3 details the full pipeline of I-GWAS' framework for privacy-preserving releases of interdependent GWASes. From lines 6 to 13, I-GWAS check if there exists a safe batch of requests for a single-GWAS. If this is the case, those requests are checked now

**Algorithm 3** Full I-GWAS workflow.

```
1:  procedure I-GWAS WORKFLOW FOR INTERDEPENDENT GWASES(G)
2:     Input: set G of GWASes
3:     Output: updated statistics of a safe GWAS g
4:     Uses: $NtoL_{single}(g)$ returns the list of selected biocenters
        and their corresponding batch of requests to update a single-GWAS g; and
        $NtoL_{I\text{-}GWAS}(g, G, Add\_Req, Rmv\_Req)$ returns the list of selected biocenters
        and their corresponding batch of requests to update GWAS g
5:     isSafeSingleBiocList := ∅
6:     isSafeFinal := False
7:     selected_SNPs := ∅
8:     for g in G do //check each existing GWAS g
9:        isSafeSingleBiocList := $LtoN_{single}(g)$
10:       if (isSafeSingleBiocList ≠ ∅) then // assemble the requests from selected biocenters
11:          for b in isSafeSingleBiocList do
12:             Add_Reqs := isSafeSingleBiocList.addRequests
13:             Rmv_Reqs := isSafeSingleBiocList.rmvRequests
14:          end for
15:       end if
16:       isSafeFinal = $isSafe_{I\text{-}GWAS}(g, G, Add\_Reqs, Rmv\_Reqs)$
17:       if (isSafeFinal) then // update the requests from selected biocenters
18:          selected_SNPs := $checkInterdependentMembership(Add\_Reqs, Rmv\_Reqs, G)$
19:       end if
20:    end for
21:    $computeTestStats$(Add_Reqs, Rmv_Reqs) // update test statistics over selected requests
22:    $computeAggregateStats$(selected_SNPs) // update aggregate statistics over selected SNPs
23: end procedure
```

considering interdependent GWASes (line 14, which is the combination of Algorithms 1 and 2). If this evaluation succeeds, aggregate GWAS statistics can be computed over the selected genomes and SNPs and are publicly published. Only the SNPs identified as safe by Algorithm 2 will have both aggregate and test statistics released. On the other hand, the others (unsafe) SNPs are secluded from having their GWAS statistics released. Note that only the I-GWAS's TEE enclave has access to the identified unsafe SNP ids and their respective GWAS statistics, and therefore these SNPs cannot be leveraged by adversaries to mount genomic privacy attacks.

## ACKNOWLEDGMENTS

We would like to express our gratitude to the anonymous reviewers for devoting the necessary time and effort to review the manuscript. We sincerely appreciate the useful comments and insightful suggestions that allowed us to improve the quality of the manuscript.

**Funding source.** This work was supported by the European Union under the H2020 Programme Grant Agreement No. 830929 (CyberSec4Europe).